\documentclass[aps,twocolumn,showpacs]{revtex4}
\usepackage{graphicx}
\usepackage{epsfig,color}
\usepackage{bm}

\begin{document}
%\DeclareGraphicsRule{.jpg}{eps}(.jpg.bb){'jpeg2ps -r 0 -h #1}
%\tighten
%\draft
%\preprint{
%\vbox{
%\hbox{\today}
%\hbox{Pusan}
%}}

%\preprint{
%\vbox{
%\hbox{\today}
%\hbox{Tashkent}
%}}
%%%%%%%%%%%%%%%%%%%%%%%%%%%%%%%%%%%%%%%%%%%%%%%%%%%%%%%
\setcounter{page}{1}
\newcommand{\re}[1]{(\ref{#1})}
\newcommand{\lab}[1]{\label{#1}}
\newcommand{\ci}[1]{\cite{#1}}
\renewcommand{\baselinestretch}{1.25}
\newcommand{\bfr}{\begin{flushright}}
\newcommand{\bfl}{\begin{flushleft}}
\newcommand{\efl}{\end{flushleft}}
\newcommand{\efr}{\end{flushright}}
\newcommand{\bc}{\begin{center}}
\newcommand{\ec}{\end{center}}
\newcommand{\be}{\begin{equation}}
\newcommand{\ee}{\end{equation}}
\newcommand{\bea}{\begin{eqnarray}}
\newcommand{\eea}{\end{eqnarray}}
\newcommand{\ba}{\begin{array}}
\newcommand{\ea}{\end{array}}
\newcommand{\nn}{\nonumber}
\newcommand{\edc}{\end{document}}
\newcommand{\ul}{\underline}
\newcommand{\ri}{\rightarrow\infty}
\newcommand{\li}{\leftarrow\infty}
\newcommand{\ra}{\rightarrow}
\newcommand{\la}{\leftarrow}
\newcommand{\ds}{\displaystyle}
\newcommand{\dsf}{\displaystyle\frac}
\newcommand{\dt}{\Delta{t}}
\newcommand{\il}{\int\limits}
\newcommand{\pal}{\partial}
\newcommand{\xxx}{{\it{X}}}
\newcommand{\bone}{{\bf 1}}
\newcommand{\gComment}[1]{}
\renewcommand{\gComment}[1]{\textcolor{red}{Gerardo: #1}}

%\begin{document}
\title[]{
Anyon Bosonization of $2D$ Fermions and Single Boson Phase Diagram
Implied from Experiment on Visualizing Pair Formation in
Superconductor $Bi_2Sr_2CaCu_2O_{8+\delta}$ }

%\author{B. Abdullaev$^1$}
%\author{C.-H. Park$^1$}
%\author{K.-S. Park$^2$}

%\author{B. Abdullaev, C.-H. Park, and K.-S. Park }
\author{B. \surname{Abdullaev$^{1,2}$}}
%\author{U. \surname{R\"{o}ssler$^2$}}
\author{C. -H. \surname{Park$^2$}}
\author{M. M. \surname{Musakhanov$^{1}$}}

\affiliation{
%\address{$^1$Institute of Applied
 Physics, National University of
Uzbekistan, Tashkent 100174, Uzbekistan.\\
%$^2$Institute for Theoretical Physics, University of Regensburg,
%D-93040 Regensburg, Germany\\
$^2$Research Center for Dielectric and Advanced Matter Physics,
Department of Physics,\\ Pusan National University, 30
Jangjeon-dong, Geumjeong-gu, Busan 609-735, Korea.}

\date{Received \today }

\begin{abstract}
Recently, Gomes {\it et} {\it al.}~\ci{Gomes} have visualized the
gap formation in nanoscale regions (NRs) above the critical
temperature $T_c$ in the high-$T_c$ superconductor
$Bi_2Sr_2CaCu_2O_{8+\delta}$. It has been found that, as
the temperature lowers, the NRs expand in the bulk superconducting state
consisted of inhomogeneities. The fact that the size of the
inhomogeneity~\ci{Pan} is close to the minimal size of the
NR~\ci{Gomes} leads to a conclusion that the superconducting phase
is a result of these overlapped NRs. In the present paper we
perform the charge and percolation regime analysis of NRs and show
that at the first critical doping $x_{c1}$, when the
superconductivity starts on, each NR carries the positive electric
charge one in units of electron charge, thus we attribute the NR
to a single hole boson, and the percolation lines connecting these
bosons emerge. At the second critical doping $x_{c2}$, when the
superconductivity disappears, our analysis demonstrates that the
charge of each NR equals two. The origin of $x_{c2}$ can be
understood by introducing additional normal phase hole fermions in
NRs, whose concentration appearing above $x_{c1}$ increases
smoothly with the doping and breaks the percolation lines of
bosons at $x_{c2}$. The last one results in disappearing the bulk
bosonic property of the pseudogap (PG) region, which explains the
upper bound for existence of vortices in Nernst effect~\ci{Wang}.
Since~\ci{Gomes} has demonstrated the absence of NRs at the PG
boundary one can conclude that along this boundary, as well as in
$x_{c2}$, all bosons disappear. As justification of appearance of
single bosons, the bosonization of $2D$ fermions is rigorously
proven using the concept of anyons. The linear density dependence
of the energy gap between excited fermionic and bosonic ground
states describes the Uemura relation for $2D$ superconductors.
\end{abstract}

%\keywords{anyon, ground state, strong magnetic field.}
%\pacs{71.10.Pm,\, 71.10.Ca,\, 71.10.Hf,\, 73.43.Cd }
%\pacs{Pacs Numbers: XXXXX}

\maketitle

\newpage

\section{Introduction}
\label{sec1}

The origin of PG and high-temperature superconductivity phases in
copper oxides is the most puzzling and challenging problem in
condensed matter physics. Despite on the intensive experimental
and theoretical studies we have no clear understanding of these
phases so far. A relationship between two phases has become a
subject of wide range theoretical proposals and their possible
experimental testing. A precursor scenario for the PG state
supposes pairing correlations without superconducting phase
coherence \ci{Emery}. This scenario has been confirmed by
experiments~\ci{Wang,Corson}. A description for the PG phase based
on the electronic competing order mechanism with experimental
arguments was given in~\ci{Tallon}. Other observations have
associated the PG with a real space electronic
organization~\ci{Vershinin} which is dominant at low dopings.

%%%%%%  one intoduces correction...
The fundamental property of the PG is a partial gap in the density
of states~\ci{Timusk} which is observed in various experiments.
%%%%%% we added "real space" ....
To understand the nature of this gap the real space atomic scale
scanning tunneling microscopy measurements of the copper oxide
$Bi_2Sr_2CaCu_2O_{8+\delta}$ have been performed. For the case of high-$T_c$
superconductivity the spatial gap inhomogeneities have been observed
in~\ci{Howald,McElroy}, while Pan {\it et} {\it al.}~\ci{Pan}
explicitly determine their minimal size. The evolution of the
%%%% we have added "a nanoscale" ....
nanoscale gap formation with temperature decrease in the PG region
has been investigated by Gomes {\it et} {\it al.}~\ci{Gomes}.

%%%% we have rewritten  and divided the sentences:
In the present paper we study the origin of minimal size NRs,
which were visualized in Refs.~\ci{Gomes,Pan} through the
measurement of the energy gap. We use the experimental fact that
PG and superconductivity phases are formed from the NRs.
Particularly, we are interested in the electric charge of NRs. We
will employ the information about the charge to understand some
ingredients of doping-temperature phase diagram of
$Bi_2Sr_2CaCu_2O_{8+\delta}$ copper oxide. The generalization of
our consideration to other cuprates will be given as well. It is
worth to notice that all physical findings in the paper are
inferred from the analysis of data for the NRs in~\ci{Gomes,Pan}.
The most important fermionic nature of the second hole inside NR
at $x_{c2}$ and dopings below $x_{c2}$ is implied from the meaning
of the second critical doping $x_{c2}$: at this doping the
superconductivity and hence, the bosonic property of the matter
disappears. The justification of the appearance of single hole
bosons will be given using the concept of anyons. In such a
treatment the anyon vector potential and the corresponding
statistical magnetic field represent Berry connection and
fictitious magnetic field~\ci{koizumi1}, respectively. A
significant role of the fictitious magnetic field, as a real
quantity originated from the Berry phase, has been stressed in
Refs.~\ci{koizumi2,koizumi3} in the non-pairing mechanism of
high-$T_c$ superconductivity. We will demonstrate that $2D$
fermions can be bosonized. So that the fermion ground state
becomes an excited state with respect to the boson one. The linear
density dependence of the energy gap between these two states
describes the well-known Uemura relation for $2D$ superconductors.

In Sec. \ref{sec2} we describe the charge and percolation
analysis of NRs on the base of experimental data given in
Refs.~\ci{Gomes} and~\ci{Pan}. The analysis provides the
interpretation of some elements of the phase diagram
doping-temperature in $Bi_2Sr_2CaCu_2O_{8+\delta}$ compound.
Sec. \ref{sec3} is devoted to a
rigorous proof of the bosonization of $2D$ fermions and existence
of the bosonic
ground state for any $2D$ quantum system. Another implication of the
bosonization to investigation of the universal
Uemura relation for $2D$ superconductors is outlined in Sec.
\ref{sec4}. We summarize and conclude our paper in Sec.
\ref{sec5}.

\section{Experiment Implied Single Boson Phase Diagram}
\label{sec2}

The authors of Ref.~\ci{Gomes} have visualized the NRs in the PG
region of $Bi_2Sr_2CaCu_2O_{8+\delta}$ compound at fixed hole
dopings $x=0.12,0.14,0.16,0.19,0.22$. It has been determined that
for $x=0.16$ and $x=0.22$ the minimal size of the NRs is
$\xi_{coh}\approx 1-3$ nm. The estimated minimal size of NRs,
$\xi_{coh}$, is about 1.3 nm in the superconducting phase~\ci{Pan}
($T_c=84 K$). Another notable result obtained in Ref. \ci{Pan} is
the observation of spatial localization of the dopped charges. The
charges are localized in the same area as NRs \ci{Pan} with the
same coherence length $\xi_{coh}$. Below we will demonstrate that
the next spatial parameter, the mean distance between two holes,
$r_0$, is important to understand the underlying physics. The
experimental doping dependence of $r_0$ can be approximated by the
relationship $r_0\approx a/x^{1/2}$ (see Fig. 34 in
Ref.~\ci{Kastner}), where $a$ is a lattice constant in the
elementary structural plaquette for the $CuO_2$ $a - b$ plane of a
copper oxide. This relationship is derived in Ref.~\ci{Kastner}
for $La_{2-x}Sr_xCuO_4$ compound with $a\approx 3.8 \AA$. It
is valid for our compound as well since the lattice constant $a$
of $Bi_2Sr_2CaCu_2O_{8+\delta}$ is $a\approx 3.8 \AA$ (see the
capture to Fig.2 in Ref.~\ci{McElroy}). It is worth to mention
that $b\approx a$ for the lattice constant $b$
of the same structural plaquette.

A principal part of our analysis is the doping $x$ dependence of
the NR charge $(\xi_{coh}/r_0)^2$. We start with
a case of zero temperature. The parameter
$\xi_{coh}/r_0$ contains an essential information in our
consideration. The factor $(\xi_{coh}/r_0)^2$ reduces to the
expression $x(\xi_{coh}/a)2$ which has a simple physical meaning:
it is a total electric charge of $(\xi_{coh}/a)^2$ number of
plaquettes each of them having a charge $x$. On the other hand,
the parameter $\xi_{coh}/r_0$ describes the average spatial
overlapping degree of two or more holes by one NR. If
$\xi_{coh}/r_0>1$ then all NRs will be in close contact to each
other providing by this the bulk superconductivity in percolation
regime.

In the Table I we outline the doping $x$ dependencies for the
function $(\xi_{coh}/r_0)^2$ for fixed experimental values
$\xi_{coh}=10\AA$ (the minimal size of the NR) and
$\xi_{coh}\approx13\AA$ taken from Ref.~\ci{Gomes} and
Ref.~\ci{Pan}, respectively, and for the function $\xi_{coh}$
which fits $(\xi_{coh}/r_0)^2$ to $(10\AA/r_0)^2$ at
$x=0.28$ and for $x=0.05$ provides $(\xi_{coh}/r_0)^2\approx 1.0$.
Numerical values of the $\xi_{coh}/r_0$ are also shown in the table.

Since in Ref.~\ci{Gomes} every NR location in the sample is
tracked with the precision $0.1\AA$, we suppose that the
$\xi_{coh}=10\AA$ has been measured with a high enough accuracy. In
addition, we assume that in Ref.~\ci{Pan} the $\xi_{coh}\approx
13\AA$ is measured with the same accuracy at $x=0.14$. Under this
condition, we conclude that the tendency of $\xi_{coh}$ to growth
from $10\AA$ to $13\AA$ when $x$ decreases from $0.22$ to $0.14$
reflects quantitatively the underlying physics. The data for
the resulting parameter $\xi_{coh}/r_0$ is approximated by the
function $2.2x^{1/3}$. The analytic equation for
$\xi_{coh}$ expressed in terms of the lattice constant $a$ is
given by $\xi_{coh}\approx 2.2 a/x^{1/6}$.

As seen from Table I, the charges $(10\AA/r_0)^2$,
$(13\AA/r_0)^2$, and $(\xi_{coh}/r_0)^2$ vary continuously with
the doping $x$. This is not surprising because they are functions
of $r_0(x)$ and $\xi_{coh}(x)$. From the analysis at
the first critical doping, $x_{c1}=0.05$, it follows that the
charge $(\xi_{coh}/r_0)^2$ of the visualized NR in Ref.~\ci{Gomes}
equals $+1$. So that, it corresponds to the charge of a single
hole. Notice, at the critical doping $x_{c1}=0.05$ the percolation parameter
is given by $\xi_{coh}/r_0=1.0$. That means the whole sample is entirely
covered with mini areas $\xi_{coh}^2=r_0^2$ contacting each other.
It is unexpected that at the second critical doping,
$x_{c2}=0.28$, the charge of the visualized NR takes the value
$+2$. This implies that at $\xi_{coh}^2=2r_0^2$ one has a pair of
holes inside the NR and, as a result, the superconductivity
disappears completely. For $x_{c2}=0.28$ we have
$\xi_{coh}/r_0>1.0$, so that the charge conductivity of the
fermions still remains.

Notice, that there are no particles
in the nature with the fractional charge, except the
quasiparticles which can be produced by many-body correlations
like in the fractional quantum Hall effect~\ci{Laughlin}. Hence,
the problem of the presence of the extra fractional charge inside
the NR has to be solved yet.
We remind~\ci{Gomes,Pan} that PG visualized NRs
constitute the bulk superconductivity phase below the critical
temperature $T_c$, and therefore, they are a precursor for that
phase. This implies undoubtedly that the NRs represent bosons at
least. At $x_{c1}=0.05$ one has the charge $(\xi_{coh}/r_0)^2=1$,
so that one may conjecture that the NR represents just a boson
localized in the square box $\xi_{coh}^2$.

For $x>0.05$ the charge $(\xi_{coh}/r_0)^2$ has an additional to
$+1$ fractional part. We assign the last one to the fractional
part of the charge of fermion. Thus the total charge
$(\xi_{coh}/r_0)^2$ of the NR includes the charge $+1$ of the
boson and the fractional charge of the fermion. However, as it was
mentioned above, the fractional charge can not exist. Therefore,
we take the number $N_{ob}$ of NRs to be equaled to the inverse
value of the fractional part to form a charge $+1$ of the fermion.
As a result, we obtain one fermion surrounding by $N_{ob}$ bosons.
The values of $N_{ob}$ are outlined in the last column of the
Table 1.

The NRs introduced in such a manner allow to understand clearly
the evolution of the fermions in the whole range $0.05 \leq x \leq
0.28$ of doping and to explain the origin of the second critical
doping $x_{c2}=0.28$. It is clear, as $x$ increases, the number of
fermions grows up inside the superconducting phase. By this, at
$x_{c2}$, when the number of fermions becomes equal to the number
of bosons, one has the breaking of the boson percolation lines,
and, thus the superconductivity disappears.

It is worthwhile to compare $\xi_{coh}$ with the lattice constant
$a$ of $Bi_2Sr_2CaCu_2O_{8+\delta}$ compound when the doping $x$
varies. We have $2.6a\leq \xi_{coh}\leq 4.5a$ for variation of $x$
from $x_{c2}$ to $x_{c1}$. However, it is well known that the
antiferromagnetic dielectric parent materials are characterized by
a strong short range magnetic interaction within the atomic length
scale $a$. Therefore, one may assume that $a$ is a length
parameter for these compounds. The fact that the size $\xi_{coh}$
is larger than $2.6a$  leads to a conclusion that the visualized
NRs are independent from the dielectric environment (the latter
forms only the spatial square shape of the NR). Due to this, the
numerical values $x_{c1}=0.05$ and $x_{c2}=0.28$ are universal for
all hole doped cuprates. However, at the second critical doping
$x_{c2}$ the length scale of boson and fermion (the half of
$\xi_{coh}$) inside NR is comparable to $a$. Therefore, the
parent compound starts to play a role from the critical doping
$x_{c2}$. Furthermore, since a coincidence of $x_{c2}$ with the PG
boundary at a zero temperature has been observed in various
experiments and for all temperatures of this boundary no NRs,
which exhibit gaps, were detected~\ci{Gomes}, the plausible
intuitive finding would be the total disappearance of bosons along
the PG bound line. So that two fundamental phenomena -- the
breaking of the boson percolation lines and the disappearance of
bosons -- occur at $x_{c2}$. The first phenomenon indicates the
end of the bulk bosonic property and the end of the $T_c$ curve as
well, whereas the second phenomenon corresponds to the end of the
bosonic property in general.  For the PG region the disappearance
of the bulk bosonic property was detected by observing the onset
temperature, $T_{onset}$, for the existence of vortices in the
Nernst effect~\ci{Wang}. The vortices have been seen so far only
in quantum Bose systems. Further evolution of fluctuations with
temperature increase destroys the bosons which totally vanish at
PG boundary.

\begin{table}[tb]
\begin {center}
\begin{tabular}{|c|c|c|c|c|c|c|} \hline
    $x$   & $(10 \AA/r_0)^2$ & $(13 \AA/r_0)^2$ & $\xi_{coh}(\AA)$ & $(\xi_{coh}/r_0)^2$ &
$\xi_{coh}/r_0$   & $N_{ob}$ \\ \hline
    0.28  &    1.939         &     3.277        &      10          &   1.939
& 1.393 &  $\sim 1$ \\ \hline
    0.22      &  1.524       &     2.575        &      10         &    1.524
& 1.235 &  $\sim 2$ \\ \hline
    0.16      &  1.108       &     1.873        &      11         &    1.341
& 1.158 &  $\sim 3$ \\ \hline
  0.14      &    0.969       &     1.638        &      12         &    1.396
& 1.182 &  $\sim 3$ \\ \hline
  0.10      &    0.693       &     1.170        &      13         &    1.170
& 1.082 &  $\sim 6$ \\ \hline
  0.05      &    0.346       &     0.585       &       17         &    1.000
& 1.000 &  \\ \hline
  0.04      &    0.277       &     0.468       &       18         &    0.897
& 0.947 &  \\ \hline
  0.02      &    0.139       &     0.234       &       20         &    0.554
& 0.744 &  \\ \hline
\end{tabular}
\end{center}
\vskip -.5cm \caption{The doping $x$ dependencies of NR charges. The
doping $x$ dependencies for $(10 \AA/r_0)^2$, $(13 \AA/r_0)^2$ at
fixed $\xi_{coh}=10 \AA$ and $\xi_{coh}=13 \AA$, respectively, for
the coherent length $\xi_{coh}$, the charge $(\xi_{coh}/r_0)^2$
and the percolation parameter $\xi_{coh}/r_0$ at this $\xi_{coh}$
are presented. The values for the number $N_{ob}$ of bosons
surrounding every fermion are shown in the last column.} \vskip
-.5cm \lab{tab-1}
\end{table}

%%%%%%%%%%%%%%%%%%%%%%fig1%%%%%%%%%%%%%%%%%%%%%%%%%%%%%%%%%%%%%%%
\begin{figure}
\begin{center}
%\DeclareGraphicsRule{.jpg}{eps}(.jpg.bb){'jpeg2ps -r 0 -h #1}
%\includegraphics[width=8cm,%
%type=eps,read=.bb,command='jpeg2ps -r 0 -h #1]{phasediagr.jpg}
%\includegraphics[bb=-100 -100 100 100]{phasediagr.jpg}
%\includegraphics[angle=0,width=9cm,scale=1,bb=-100 -100 100 100]{fig11.ps}
%\includegraphics[width=9cm,scale=1]{gr2.eps}
\includegraphics[width=8cm,scale=1]{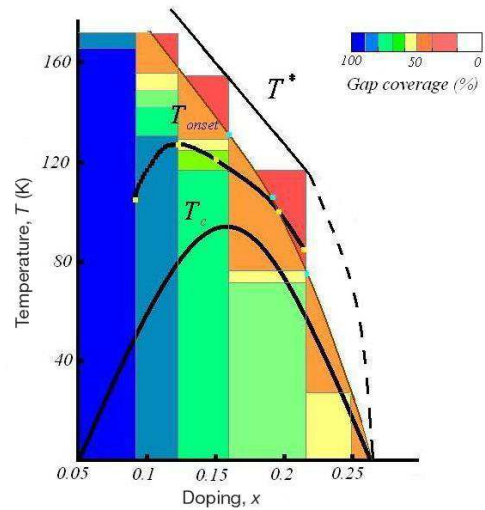}
\end{center}
\caption{Schematic single hole bosonic phase diagram for $Bi_2Sr_2CaCu_2O_{8+\delta}$.} \lab{fig1}
\end{figure}
%%%%%%%%%%%%%%%%%%%%%%fig1%%%%%%%%%%%%%%%%%%%%%%%%%%%%%%%%%%%%%%%

The schematic single hole bosonic phase diagram for
$Bi_2Sr_2CaCu_2O_{8+\delta}$ is depicted in the Fig. 1. The
coloured zones indicate the percentage of the sample that is
gapped at given temperature and doping (in analogy with the phase
diagram shown in Ref.~\ci{Gomes}). The solid lines correspond
to the following observed temperatures: PG boundary
$T^*$ and onset temperature $T_{onset}$ for Nernst effect signals
taken from Ref.~\ci{Wang}, and the critical temperature $T_c$ from
Ref.~\ci{Gomes}. The extrapolation of the connection of $T^*$ with
the second critical doping, $x_{c2}$, is depicted by the dashed
line. The yellow points correspond to fixed $T_{onset}$ values
from Ref.~\ci{Wang}, and the blue points represent the temperature
data for $50\%$ of gapped area of the sample from Ref.~\ci{Gomes}
measured at fixed dopings. The thin brown coloured solid line fits
the blue points. The percentage for the gapped doping is
calculated by using the equation $(1-1/ (N_{ob}+1))\cdot 100\%$
under the assumption that the NRs overlap each other. It is
remarkable that $T_{onset}$ line is substantially located in the
brown coloured zones which means there is no bulk bosonic property
above these zones. It is worth to compare the homogeneous $100\%$
gap coverage observed in Ref.~\ci{Gomes} with our proposed varying
one in Fig. 1 for low temperature and doping levels $0.12\leq
x\leq 0.22$. Employing the doping changing dynamics of the $50\%$
gap coverage obtained in Ref.~\ci{Gomes}, we find that this
percentage is applicable also at the second critical doping,
$x_{c2}$, which is shown in Fig. 1 by yellow ($60\%$) and brown
($50\%$) colours. On the other hand, if we consider the NR charge
$+1$ for the above interval of doping with further its increasing
up to $+2$, close to $x_{c2}$, then we will reproduce exactly the
percentages observed in the phase diagram in Ref.~\ci{Gomes}.

The next interesting finding is that the number of external
interstitial atoms sufficient to produce one doped hole in the
dielectric parent material equals to $1/x$. For
$La_{2-x}Sr_xCuO_4$ compound it is a number of $Sr$ atoms, since
the hole doping and the concentration of atoms are expressed by
$x$. In the interval $0.05\leq x\leq 0.28$ this number varies from
$20$ to $3$.

At the end of this section, we discuss on the percolation
threshold of $2D$ classical systems and compare it with our $50\%$
one used for bulk bosonic property. If we remind white and black
cells of the chessboard and assume that the black ones represent a
region in which the percolation should occur, it becomes clear
that the percolation threshold consists of $50\%$ coverage by this
colored region of the whole chessboard area. The experiment for
some particular system indicates to more than $40\%$ coverage for
its value~\ci{Smith} (the numerical simulation for the same system
has confirmed this observed result~\ci{Weinrib}).

\section{Real Bosonization of $2D$ Fermions}
\label{sec3}

It might be seem that the appearance of the single hole bosons in
the real matter is too exotic and never can be realized. Since it
is hard even to imagine that except a pairing of fermions, as in
conventional superconductors, some other mechanisms can lead to
transformation of fermions into bosons. However, the fact that $a
- b$ planes of $CuO_2$ atoms play a dominant role in the
determination of the physics of cuprates provides an opportunity
to exploit the fundamental property of the two-dimensionality.
Specifically, the $2D$ topology allows the fractional statistics
\ci{lei} characterized by a continuous parameter $\nu$ taking
values between 0 (for bosons) and 1 (for fermions). The particles
with $0<\nu<1$ are generically called anyons \ci{wil2,wil1}. The
quasiparticle excitations in the fractional quantum Hall regime
\ci{fis,ler,kh} and in certain quantum magnets \cite{kitaev} can
be described using the anyon concept. One can apply the last one
in the study of properties of the mentioned above $a - b$ planes.

In this section we will present the rigorous derivation of the
real bosonization of 2$D$ fermions. It can be achieved by exact
cancellation of terms in the ground state energy arisen from
fermion (anyon) statistics and a Zeeman interaction of spins
$\hbar /2$ of particles with statistical magnetic field
\ci{ler,Dunne} produced by vector potential of anyons. As in our
recent papers~ \ci{aormn,aom}, the calculation will be carried out
in the framework of a variational approach. However, we do not use
the cut-off regularization procedure of the logarithmic divergence
for the nearest interparticle distance, as it was done in
\ci{aormn}, since the short-range correlations of particles will
be accurately taken into account in the system wave function. In
\ci{abdullaev1} the bosonization of 2$D$ fermions has been
obtained approximately using the cut-off regularization.

Let us consider the Hamiltonian \bea \ba{r} \hat
H=\dsf{1}{2M}\ds\sum_{k=1}^N\left[\left(\vec p_k+\vec A_{\nu}(\vec
r_k)\right)^2+ M^2\omega_0^2 |\vec{r_k}|^2 \right] \\
+ \dsf{1}{2}\ds\sum_{k=1}^N\left[V(\vec r_k) +
\ds\sum_{j\not=k}^N\dsf{e^2} {|\vec r_{kj}|}\right] \
\lab{gsetup1} \ea \eea of the gas of $N$ anyons with mass $M$ and
charge $e$, confined in $2D$ parabolic well, interacting through
Coulomb repulsion potential in the presence of uniform positive
background~\ci{Laughlin} $ V(\vec r_k)$. Here, $\vec r_k$ and
$\vec p_k$ represent the position and momentum operators of the
$k$th anyon in 2$D$ space dimension, \be \vec A_{\nu }(\vec
r_k)=\hbar\nu\ds\sum_{j\not=k}^N\dsf{\vec e_z \times\vec r_{kj}}
{|\vec r_{kj}|^2} \lab{gsetup2} \ee is the anyon gauge vector
potential \ci{wu}, $\vec r_{kj}=\vec r_k-\vec r_j$, and $\vec e_z$
is the unit vector normal to the 2$D$ plane. In the expression for
$\vec A_{\nu }(\vec r_k)$ and hereafter we assume that $0\leq \nu
\leq 1$.

In the bosonic representation of anyons we take the system wave
function in the form~\ci{Comtet} (see also \ci{ouvgroup}): \be
\Psi(\vec R)=\prod_{i\not=j}r_{ij}^{\nu}\Psi_T(\vec R).
\lab{gsetup7} \ee Here $\vec R = \{\vec r_1....\vec r_N\}$ is the
configuration space of the $N$ anyons. The product in the right hand
side of this equation is the Jastrow-type wave function. It
describes the short distance correlations between two particles due
to anyonic (fermionic) statistics interaction.

%Let us check the quality of $\Psi(\vec R)$ for the entire range of
%the anyon parameter $\nu$ on the example of harmonically confined
%anyons without Coulomb interaction.
Let us consider first the term in the Hamiltonian $\hat H$,
Eq.~\re{gsetup1} containing the anyon vector potential $\vec
A_{\nu }(\vec r_k)$. Substituting $\Psi(\vec R)$,
Eq.~\re{gsetup7}, in Schredinger equation with this Hamiltonian,
we obtain an equation $\widetilde{\hat H}\Psi_T(\vec
R)=E\Psi_T(\vec R)$ with the novel Hamiltonian $\widetilde{\hat
H}=\widetilde{\hat H}_1+\widetilde{\hat H}_2$, where \be
\widetilde{\hat
H}_1=\ds\sum_{k=1}^N\left(\dsf{-\hbar^2\Delta_k}{2M}-
\dsf{\hbar^2\nu}{M}\ds\sum_{j\not=k}\dsf{{\vec r}_{kj}\cdot {\vec
\nabla}_k}{|\vec r_{kj}|^2}\right) \lab{gsetup7b} \ee
%\\+ \dsf{1}{2}\ds\sum_{k=1}^N\left(V(\vec r_k) +
%\ds\sum_{j\not=k}^N\dsf{e^2} {|\vec r_{kj}|}\right) \lab{gsetup7b}
%\ea \eea
and \be \widetilde{\hat
H}_2=-i\dsf{\hbar}{M}\ds\sum_{k=1}^N\left(\vec A_{\nu }(\vec
r_k)\cdot {\vec \nabla}_k+\nu \ds\sum_{j\not=k}\dsf{\vec A_{\nu
}(\vec r_k)\cdot {\vec r}_{kj}}{|\vec r_{kj}|^2}\right).
\lab{gsetup7c} \ee

As shown in Ref.~\ci{Comtet}, the $\nu$ interaction Hamiltonian in
$\widetilde{\hat H}_1$, i.e., the second its term, is equivalent
to a sum of two-body potentials \be \dsf{\pi\hbar^2 \nu
}{M}\ds\sum_{j\not =k} \delta ^{(2)}( \vec r_k- \vec r_j) \ .
\lab{gsetup7aa} \ee Therefore, the Hamiltonian $\widetilde{\hat
H}_1$ now reads\be \widetilde{\hat
H}_1=\ds\sum_{k=1}^N\left(\dsf{-\hbar^2\Delta_k}{2M}+
\dsf{\pi\hbar^2 \nu }{M}\ds\sum_{j\not =k} \delta ^{(2)}( \vec
r_k- \vec r_j)\right). \lab{gsetup7ab} \ee

The wave function $\Psi(\vec R)$ has been used in Refs.~\ci{Comtet}
and \ci{ouvgroup} in approximate perturbative treatment for the
calculation of the ground state energy close to boson end of anyons
$\nu\rightarrow 0$. Being calculated without the Jastrow product
constituent, this energy displays the logarithmic divergence problem
in regard to the nearest interparticle distance on which we already
mentioned above.

To check the quality of $\Psi(\vec R)$ for the entire range of the
anyon parameter $\nu$ we consider the system of harmonically
confined anyons without Coulomb interaction and calculate its ground
state energy. To do this we add the parabolic potential $M\omega_0^2
|\vec{r_k}|^2/2$ inside the brackets in Eq.~\re{gsetup7ab} thus
redefining the Hamiltonian $\widetilde{\hat H}_1$.

In the variational scheme \ci{aormn} we minimize the expression
\be
 E=\dsf{\int \Psi_T^*(\vec R)\widetilde{\hat H}\Psi_T(\vec R) \ d\vec R}{\int
\Psi_T^*(\vec R) \Psi_T(\vec R) \ d\vec R} \ . \lab{gsetup3} \ee For
energies expressed in units of $\hbar\omega_0=\hbar^2/(ML^2)$ and
lengths in units of $L$ the normalized trial wave function has the
following form \be \Psi_T(\vec
R)=\left(\dsf{\alpha}{\pi}\right)^{N/2}\prod_{k=1}^N
\exp\left(-\alpha\dsf{(x_k^2+y_k^2)}{2}\right). \lab{gsetup7d} \ee
Here, $\alpha$ is the variational parameter.

The simple calculation shows that the expectation value $E$ with the
Hamiltonian $\widetilde{\hat H}_2$, Eq.~\re{gsetup7c}, and wave
function given by Eq.~\re{gsetup7d}, equals zero. Therefore, we will
assume in Eq.~\re{gsetup3} $\widetilde{\hat H}=\widetilde{\hat
H}_1$.

A minimization of the energy $E$ with respect to $\alpha$ gives
the expression for the ground state energy \be E_0=\hbar \omega_0
N(1+\nu(N-1))^{1/2} , \lab{gsetup7ad} \ee which coincides exactly
with Eq. (21) in our paper~\ci{aormn} found by using the cut-off
regularization. In Ref.~\ci{aormn} we have compared the exact
numeric values for the ground state energy of fermions in the
harmonic potential obtained by using the Pauli exclusion principle
with ones calculated by using the Eq.~\re{gsetup7ad} for $\nu=1$.
As demonstrated in Figs. 1 and 2 of~\ci{aormn}, the maximal
deviation (no more than 12$\%$) occurs at small numbers of $N$ and
this deviation tends to zero for increasing $N$. For large $N$ and
arbitrary $\nu$ the formula~\re{gsetup7ad} is consistent (up to a
numerical factor) with the approximate expression $E \approx \hbar
\omega_0\nu ^{1/2}N^{3/2}$ of Chitra and Sen \ci{chitra}
calculated in Thomas-Fermi approximation for $\nu>1/N$. It is
obvious that Eq.~\re{gsetup7ad} reproduces the result of Wu (see
the paper of Wu in Refs.~\ci{wu}) in the bosonic limit
$\nu\rightarrow 0$. This analysis unambiguously shows that the
Hamiltonian $\widetilde{\hat H}_1$, Eq.~\re{gsetup7ab}, reproduces
accurately the anyon statistics interaction for all values of
$\nu$ and $N$ (there is no doubt that the harmonically confinement
potential does not affect on the statistics interaction).

Now we demonstrate the real bosonization of 2$D$ fermions on the
example of anyons in parabolic well. To do this we consider the
Zeeman interaction term \be \dsf{\hbar }{M}\ds\sum_{k=1}^N {\hat
{\vec s}} \cdot \vec b_k \ \lab{gsetup8} \ee of spins with the
statistical magnetic field \ci{Dunne} (see also \ci{ler}) \be \vec
b_k = -2\pi \hbar \nu \vec e_z \ds\sum_{j\not =k} \delta ^{(2)}(
\vec r_k- \vec r_j) \ , \lab{gsetup9} \ee which can be derived if
one calculates $\vec b_k = \vec \nabla \times \vec A_{\nu }(\vec
r_k)$ by using Eq.~\re{gsetup2}.

The sign in Eq.~\re{gsetup8} is taken according to the standard
definition $-\vec \mu \vec H$ of the Zeeman term \ci{landaus},
where $\vec \mu$ and $\vec H$ are a magnetic moment of a spin and
an external magnetic field, respectively. It takes into account
also self consistently the charge sign of particles and direction
of the statistical magnetic field $\vec b_k$. For electrons with
charge $e=-|e|$ adopted in this paper the condition $\nu>0$ is
correct, while for holes with charge $e=|e|$ one needs to take
$\nu=-|\nu|$ in the expression for $\vec b_k$, and to change the
sign of Eq.~\re{gsetup8} itself. In the last case one has to
replace $\nu$ with $|\nu|$ in all our formulas below.

For $s_z=\hbar /2$ and using the expression, Eq.~\re{gsetup9}, for
$\vec b_k$ one obtains
\be \dsf{\hbar }{M}\ds\sum_{k=1}^N {\hat
{\vec s}} \cdot \vec b_k=
 -\pi  \nu \dsf{\hbar^2}{M} \ds\sum_{k(j\not =k)}
\delta ^{(2)}( \vec r_k- \vec r_j) \ . \lab{gsetup10} \ee Being
added to the expression, Eq.~\re{gsetup7ab}, for the Hamiltonian
$\widetilde{\hat H}_1$, this Zeeman term cancels exactly the
second one of $\widetilde{\hat H}_1$, which is responsible for the
statistics of fermions (for $\nu=1$)  and anyons. Since the energy
of bosons is lower than one for fermions and anyons, there appears
a coupling of spin with statistical magnetic field for every
particle or bosonization of 2$D$ fermions and anyons. From this
one can conclude, if anyon concept is correct for the description
of any $2D$ quantum system, its ground state should be bosonic
with $\nu=0$, while its excited state should be fermionic
($\nu=1$) or anyonic ($0<\nu<1$) depending of the fixed value of
$\nu$.

It might seem that the subject described in this section has no
relation to real physics since the physical sense of the statistical
magnetic field is unclear or may have an artificial meaning.
However, if we express the gauge vector potential $\vec A_{\nu
}(\vec r_k)$, Eq.~\re{gsetup2}, in the form $\vec
A^{fic}=\hbar\nu\nabla_{2D} \chi$, where the expression of the
angular variable $\chi$, Eq.~(3) of Ref.~\ci{koizumi1}, is exactly
coincides with one for anyons, Eq.~(3.2.31) of book~\ci{ler}, it
becomes evident that $\vec A_{\nu }(\vec r_k)$ represents the Berry
connection and the statistical magnetic field is the fictitious
magnetic field originated from the Berry phase~\ci{koizumi1} of
anyons (see book of Wilczek~\ci{wil1}). As it was demonstrated in
Refs.~\ci{koizumi1,koizumi2,koizumi3}, the fictitious magnetic field
is a real physical quantity, and its role is significant in the
construction of the alternative mechanism of high-$T_c$
superconductivity.

At the end of this section, one can say that the bosonic ground
state nature of the arbitrary $2D$ quantum system is the intrinsic
fundamental property of the two dimensionality, which originates
from its topology. In the light of this finding, the appearance of
single bosons in the experiment of Gomes {\it et} {\it
al.}~\ci{Gomes} and discussed in the previous section  might be
not occasional. Another important qualitative issue, which leads
from a result of ~\ci{Gomes} experiment, is in the following. The
random positions in the real space of the observed pairs totally
exclude any mechanism for the pair formation. Since occasionally
positioned in this space coherent excitations (phonons, magnons or
other quasi-particles), which create pairs, are problematic, if
the system is homogenous. The last observation deduced from Gomes
{\it et} {\it al.} paper is the fundamental argument for the
justification of the single hole nature of the cuprate physics.

\section{Origin of Uemura Relation}
\label{sec4}

Now, it is widely accepted (see Ref.~\ci{zaanen}) that the Uemura
relation (UR), the linear dependence of $T_c$ on concentration of
charge carriers, originally observed in Refs.~\ci{uemura1} and
\ci{uemura2} for underdoped cuprate, bismuthate, organic,
Chevrel-phase and heavy-fermion superconductors, survives also for
extended class of other superconductors and has a fundamental
universal character. There is no doubt that this relation together
with other empirical Home's law, Ref. \ci{homes}, for cuprate and
conventional (but except the molecular, Ref. \ci{pratt})
superconductors plays an important role for the construction of
the mechanism of superconductivity in these materials and can even
be a source for discovering fundamental properties of the
underlying physics. It is worth to remind that the Home's law
relates a superfluid density (charge concentration) to the
electric conductivity of the normal state nearly above $T_c$ and
$T_c$ itself.

Recently, there was observed a deviation from the UR into sublinear
scale, in which $T_c$ had the dependence on carrier concentration
with power less than unit (more exactly with power close to 1/2),
for the particular $YBa_2Cu_3O_y$ cuprate (see Ref.~\ci{sonier} and
references therein) and in $Y_{0.8}Ca_{0.2}Ba_2Cu_3O_{7-\delta}$,
$Bi_2Sr_2CaCu_2O_{8+\delta}$ and $La_{2-x}Sr_xCuO_4$ copper oxides,
Ref.~\ci{tallon}. In the second case the deviation was more evident
for doping close to and beyond the optimal one. The additional
deviation from the linear concentration dependence of $T_c$, with
the power of concentration slightly great than unit (with power
3/2), was reported in  Ref.~\ci{pratt} for quasi-2$D$ molecular
superconductors.

However, the experimental data have indicated the prominent UR for
samples with nearly 2$D$ geometry: Ref.~\ci{rufenacht} for
ultrathin $La_{2-x}Sr_xCuO_4$ and Ref.~\ci{matthey} for very thin
$NdBa_2Cu_3O_{7-\delta}$. The problem with uncertainty of power of
the carrier concentration was resolved in the remarkable and
unprecedent on precision experimental investigation,
Ref.~\ci{hetel},  of $T_c$ as function of carrier concentration.
Authors of Ref.~\ci{hetel} have studied the power of the
dependence as function of number of $CuO_2$ atoms containing $a-b$
layers for $Y_{1-x}Ca_xBa_2Cu_3O_{7-\delta}$ cuprate. Varying this
number from 40 down to 2, they observed the changing the power of
concentration from 1/2 to unit, thus revealing the obvious
relation of UR to the two-dimensionality of the system. Motivated
by this observation, in the present section we investigate the
possible role of the fermion bosonization, which is a result of
the topology of $2D$, to the origin of UR.

In our previous work~\ci{aom} we have derived an analytic
expression for the ground state energy of the homogeneous $2D$
anyon gas with the Coulomb interaction. This was done for all
values of the statistics parameter $\nu$ and mean distance between
particles $r_0$ by flattening out the confining potential with a
simultaneous increase of the particle number $N$, but fixed areal
density, to obtain the infinite size system, i.e., the
thermodynamic limit. It has been achieved by the redefining the
strength $\omega_0$ of the harmonic potential in the Hamiltonian,
Eq.~\re{gsetup1}, such that it vanishes with increasing $N$.

Applying the relationship $r_0\approx a/x^{1/2}$, where $a\approx
3.8 \AA$ (see Sec. \ref{sec2}), we do an estimate values of $r_0$,
expressed in Bohr radius $a_B$ unit ($r_s=r_0/a_B$), corresponding
the doping interval $x_{c1}\leq x\leq  x_{c2}$. One obtains
$13.12\leq r_s\leq 32.14$. For this interval of $r_s$ we have
obtained in Ref.~\ci{aom} the expression \be {\cal
E}(\nu,r_s)=\dsf{E(\nu,r_s)}{NRy}=E_{WC}+\dsf{7\nu
E_{WC}^2}{3c_{WC}^2} \lab{gsetup11} \ee for the ground state
energy per particle of the Coulomb interacting anyon gas. Here,
$Ry$ is the Rydberg  energy unit and for large $r_s$ the ground
state energy does not depend on statistics and equals to the
energy of the classical 2$D$ Wigner crystal \ci{bm},
$E_{WC}=-c_{WC}^{2/3}/r_s$ with $c_{WC}^{2/3}=2.2122$.

Taking into account from the previous section that the excited
state of the $2D$ system is fermionic and the ground state is
bosonic,  one can write the explicit expression for an energy gap
between these two states \be \Delta(r_s)={\cal E}(\nu=1,r_s)-{\cal
E}(\nu=0,r_s)=\dsf{7 E_{WC}^2}{3c_{WC}^2} \ . \lab{gsetup12} \ee
The meaning of this expression in that to become the fermion the
boson should gain the energy $\Delta(r_s)$. Substituting in
Eq.~\re{gsetup12} the expression for $E_{WC}$ and introducing the
$2D$ density $n=1/(\pi r_0^2)$ one derives \be \Delta(n)=\dsf{7
\pi n a_B^2}{3c_{WC}^{2/3}} \ . \lab{gsetup13} \ee Since the
critical temperature $T_c$ is proportional to $\Delta(n)$, one can
conclude that the $2D$ topology driven bosonization of fermions
may explain the UR for variety superconductors, whose physics is
quasi - two dimensional.

In Ref.~\ci{abdullaev1}, using the expression of $\Delta(n)$ for
optimal doping, we have obtained the values of the maximal
temperature $T_{c,max}$ of the doping-temperature phase diagram
for hole and electron doped cuprates, which were close to
experimental ones.

\section{Conclusion}
\label{sec5}

Summarizing the paper, we have succeeded in understanding the
following constituents of the doping-temperature phase diagram of
the hole doped copper oxides: (i) the first and second critical
dopings have been a result of emergence and disappearance of the
single hole boson percolation lines, respectively; (ii) the
disappearance of the percolation lines leads to the end of the PG
bulk bosonic property or to the end of Nernst effect signals;
(iii) the fact that the PG boundary was a bound, where the single
hole bosons disappear, confirmed by Ref.~\ci{Gomes}. Our findings
are consistent with the recent observation~\ci{Gavrilkin} of the
superconducting phase consisted of the array of nanoclusters
embedded in the insulating matrix and of percolative transition to
this phase from the normal phase in $YBa_2Cu_3O_{6+ \delta}$.
Superconducting islands introduced in insulating background
have been used for the interpretation of the
superconductor-insulator transition in
$Bi_2Sr_{2-x}La_xCaCu_2O_{8+\delta}$ compound~\ci{Oh}. In a recent
paper~\ci{tahir} (see also~\ci{koizumi2}) a significant role of
the percolation of elementary structural plaquettes on universal
properties of cuprates has been established. Using $3D$
percolation mechanism the authors of Ref.~\ci{tahir} succeeded in
explanation of the  $T_c$ phase diagram, room-temperature
thermopower, neutron spin resonance, and STM incommensurability.

We have presented justification in the microscopic treatment of
the appearance of single bosons. In rigorous derivation we have
obtained the bosonization of $2D$ fermions, thus proposing the
bosonic ground state for any $2D$ quantum systems. The crucial
role in our consideration is assigned to the concept of anyons.
The UR for $2D$ superconductors has been also understood according
to the fermion bosonization approach. The boson and fermion mixing
nature of PG region, derived from experiment~\ci{Gomes}, is
consistent with our treatment and description of low temperature
non-Fermi liquid heat conductivity and entropy~\ci{Abdullaev2}.

\section{Acknowledgements}

The work is partially supported by Korean Research Foundation
(Grant KRF-2006-005-J02804).

\end{document}